\documentclass{article}
\usepackage{spconf,amsmath,graphicx}
\usepackage{spconf,amsmath,epsfig}
\usepackage{booktabs}
\usepackage{amsmath}
\usepackage{multirow}
\usepackage{graphicx}
\usepackage{cases} 
\usepackage[table,xcdraw]{xcolor}
\usepackage[misc]{ifsym}

\begin{document}
\title{MediumVC: Any-to-any voice conversion using synthetic specific-speaker speeches as intermedium features}
\name{Yewei Gu$^1$$^,$$^2$, Zhenyu Zhang$^1$$^,$$^2$, Xiaowei Yi$^1$$^,$$^2$, Xianfeng Zhao$^1$$^,$$^2$$^,$$^{\textrm{\Letter}}$}
\address{$^1$State Key Laboratory of Information Security, Institute of Information Engineering,\\
 Chinese Academy of Sciences, Beijing 100093, China \\$^2$School of Cyber Security, University of Chinese Academy of Sciences, Beijing 100049, China\\
\{guyewei,zhangzhenyu,yixiaowei,zhaoxianfeng\}@iie.ac.cn}


%
\maketitle
\begin{abstract}
To realize any-to-any (A2A) voice conversion (VC), most  methods are to perform symmetric self-supervised reconstruction tasks (\emph{X$_{i}$} $\to$ $\hat{\emph{X$_{i}$}}$), which usually results in inefficient performances due to inadequate feature decoupling, especially for unseen speakers. We propose a two-stage reconstruction task (\emph{X$_{i}$} $\to$ $\hat{\emph{Y$_{i}$}}$ $\to$ $\hat{\emph{X$_{i}$}}$) using synthetic specific-speaker speeches as intermedium features, where A2A VC is divided into two stages: any-to-one (A2O) and one-to-Any (O2A). In the A2O stage, we propose a new A2O method: SingleVC, by employing a noval data augment strategy(pitch-shifted and duration-remained, PSDR) to accomplish \emph{X$_{i}$} $\to$ $\hat{\emph{Y$_{i}$}}$. In the O2A stage, MediumVC is proposed based on pre-trained SingleVC to conduct $\hat{\emph{Y$_{i}$}}$ $\to$ $\hat{\emph{X$_{i}$}}$. Through such asymmetrical reconstruction tasks (\emph{X$_{i}$} $\to$ $\hat{\emph{Y$_{i}$}}$ in SingleVC and $\hat{\emph{Y$_{i}$}}$ $\to$ $\hat{\emph{X$_{i}$}}$ in MediumVC), the models are to capture robust disentangled features purposefully. Experiments indicate MediumVC can enhance the similarity of converted speeches while maintaining a high degree of naturalness.
\end{abstract}
\begin{keywords}
voice conversion, any-to-any, any-to-one, time-shifted and duration-remained
\end{keywords}
\section{Introduction}
\label{sec:intro}
Voice conversion (VC) aims to transform latent speaker-related information of source speeches to that of the target while preserving the linguistic content. Conventional statistical modeling techniques,  \emph{e.g.}, frequency warped-based\cite{shuang2006frequency} and GMM-based model\cite{toda2005spectral}, are considered to be limited by deficient feature representation. Technological advancements from statistical modeling to deep learning have made remarkable progress, especially researches towards the non-parallel corpus. Among the methods, there exist two main patterns: encoder-decoder-based and GAN-based. The former usually performs speaker-content disentanglement, then conducts the conversion through combining source speaker-independent features with target speaker-related features. The pattern can be further divided into PPG-based, ASR-TTS, and Auto-encoder \emph{etc}\cite{zhao2020voice}. Typical applications including FragmentVC\cite{lin2021fragmentvc}, AutoVC\cite{qian2019autovc}  and AdaIn-VC\cite{chou2019one},  are devoted to solving any-to-any (A2A) problems. The GAN-based models, \emph{e.g.}, CycleGAN-VC\cite{kaneko2018cyclegan,kaneko2019cyclegan,kaneko2020cyclegan}, StarGAN-VC\cite{kameoka2018stargan,kaneko2019stargan}, are designed for a predefined set of speakers, and due to lack of explicit speaker embeddings, can hardly realize A2A VC.

Learning from mel-spectrograms of source speeches,  traditional pitch-edited speeches and converted speeches, we consider content structures and prosodic information determine distributions of spectral energies along the time axis, remaining consistent among the three. Moreover, speaker-related features, \emph{e.g.}, fundamental frequency(F0) and harmonics, generally determine the distributions along the frequency axis, especially which present individual periodic patterns. Therefore, we consider  A2A tasks are to build the periodic patterns for target speakers while maintaining the distributions along the time axis. Then we propose a two-stage process to perform A2A: any-to-one (A2O) and one-to-any (O2A), meaning using the converted specific-speaker speeches as the intermedium features(SSIF) instead of traditional encoded content information.

Here we propose MediumVC, an utterance-level method towards A2A. Before that, we propose SingleVC to perform an A2O task(\emph{X$_{i}$} $\to$ $\hat{\emph{Y$_{i}$}}$, \emph{X$_{i}$} means utterance \emph{i} spoken by \emph{X}). The $\hat{\emph{Y$_{i}$}}$ are considered as SSIF. To build SingleVC, a data augment strategy: pitch-shifted and duration-remained(PSDR) is employed to produce paired asymmetrical training data. Based on pre-trained SingleVC, MediumVC performs the $\hat{\emph{Y$_{i}$}}$ $\to$ $\hat{\emph{X$_{i}$}}$ task. The asymmetrical reconstruction mode 
makes SingleVC and MediumVC advance in feature decoupling and fusion. Experiments demonstrate MediumVC performs robustly to unseen speakers across multiple public datasets.

\begin{figure}[t]
	\centering
	\includegraphics[width=1.0\linewidth]{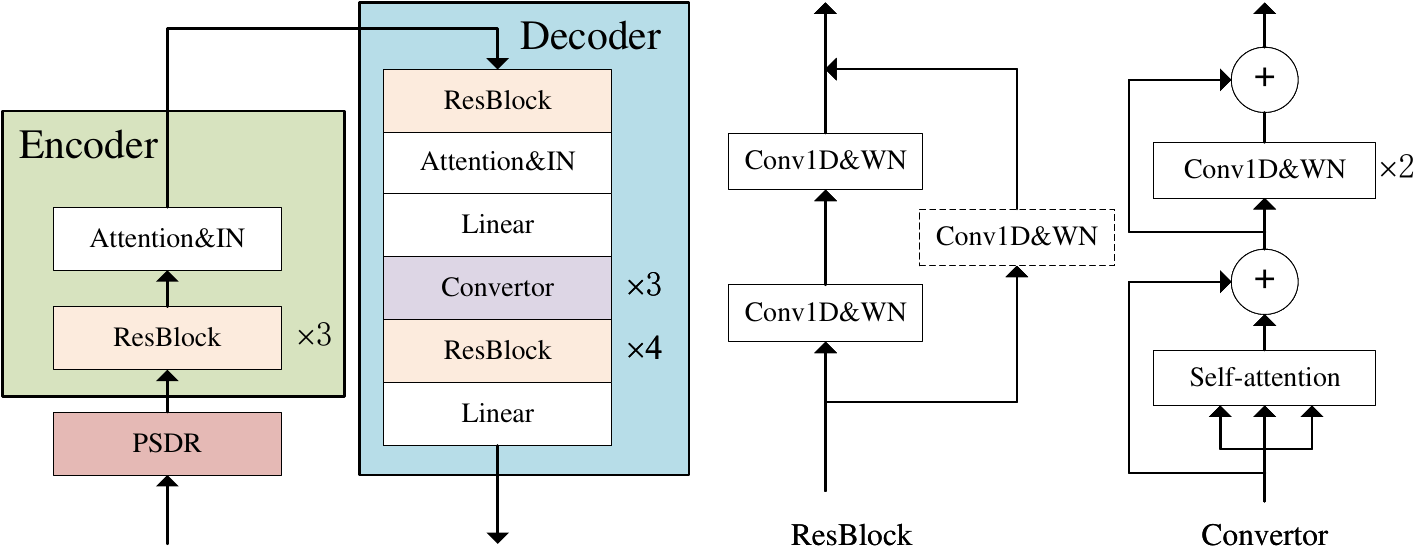}
	\caption{The architecture of SingleVC. IN means Instance Normalization and WN means Weight Normalization.}
	\label{fig:match}
\end{figure}

\section{Methods}
\label{sec:format}
\subsection{PSDR} The pitch of speeches corresponds to a set of frequencies that make up the speeches. PSDR means scaling F0 and correlative harmonics with duration remained, which intuitively modifying the speaker-related information while maintaining linguistic content and prosodic information. The frequency $f$ transformation relation is shown in Eq. 1, and the number of semitones for shifting is noted as $s$.
\begin{eqnarray}
f_{final}=2^{s/12}*f_{init}
\end{eqnarray}\\
PSDR can be used as a data augment strategy for VC by producing fake parallel corpus. To verify its feasibility that slight pitch shifts don't affect content information,  we measure the word error rate(WER) between source speeches and pitch-shifted speeches through Wav2Vec 2.0\cite{baevski2020wav2vec}-based ASR System\footnote{https://github.com/huggingface/transformers}. The speeches of p249 (female) from VCTK Corups\cite{veaux2016superseded} is selected, and pyrubberband\footnote{https://github.com/bmcfee/pyrubberband} is utilized to  execute PSDR. Table 1 indicates that when \emph{s} in -6\textasciitilde4, the strategy applies to VC with acceptable WERs.
\begin{table}[t]
	\centering
	\caption{WERs on PSDR with pitch-shifted $s$.}
	\label{tab:my-table0}
	\setlength{\tabcolsep}{1.5mm}{
		\begin{tabular}{cccccccc}
			\hline
			\textbf{S}   & -7 & -6 & -5 & 0 & 3 & 4 & 5\\ \hline
			\textbf{WER(\%)}      & 40.51  & 25.79  & 17.25  & 0  & 17.27 & 25.21 & 48.14   \\ \hline
		\end{tabular}
	}
\end{table}

\subsection{SingleVC} SingleVC performs A2O VC through a self-supervised task (\emph{X$_{i}$} $\to$ $\hat{\emph{X$_{i}^{s}$}}$ $\to$ $\hat{\emph{X$_{i}$}}$).  $\hat{\emph{X$_{i}^{s}$}}$ is  a PSDR-processed speech with pitch-shifted $s$. The architecture is shown in Fig. 1. The encoder with Instance Normalization(IN)\cite{ulyanov2016instance} is used to extract speaker-independent features into 36-dim bottleneck representations. The decoder is made of  \emph{Linear}s,  \emph{Convertor}s and  \emph{ResBlock}s. The  \emph{Convertor} is employed to capture the remote dependence  between F0 and harmonics, and converts  the correlations from pitch-shifted speeches to that of the specific-speaker. The  \emph{ResBlock}s and  \emph{Linear}s are used to reconstruct spectral energy distributions and smooth background noises. In our practice, Weight Normalization(WN)\cite{salimans2016weight} compared to Batch Normalization(BN)\cite{ioffe2015batch} and Layer Normalization(LN)\cite{ba2016layer} can prominently  improve the robustness of model. The pipeline of PSDR ($\mathcal{F}$($\cdot$,$\cdot$)), encode (\emph{E$_{v}$}($\cdot$)) and decode (\emph{D$_{v}$}($\cdot$)) are noted as Eq. 2,3, and to  minimize the  reconstruction loss as Eq. 4.
\begin{align}
\hat{X}_{i}^{s}&=\mathcal{F} (X_{i},s) \\
\hat{X_{i}}&=D_{v}(E_{v}(\hat{X}_{i}^{s})) \\
\begin{matrix}
\mathit{\mathbf{min}}\\ 
E_{v}(\cdot ),D_{v}(\cdot)
\end{matrix}&=E[\begin{Vmatrix}
X_{i}-\hat{X}_{i}^{s}
\end{Vmatrix}_{1}^{1}]
\end{align}\\
We select p249 (female, 22.5-minute) from VCTK corpus as the single corpus. Compared to males, the periodic patterns of females perform more stable due to the higher frequency resolution. To verify the robustness, SingleVC is evaluated with multiple unseen speakers from various databases. Details are presented in Table 2, and the assessment scheme is the same as Section 4.1. Results indicate the pre-trained SingleVC performs well enough to be used for downstream A2A tasks. The samples\footnote{https://brightgu.github.io/SingleVC/} and codes\footnote{https://github.com/BrightGu/SingleVC} are released online.
\begin{table}[t]
	\centering
	\caption{Evaluation On SingleVC. M,F represent male, female respectively. Each speaker provides 5 utterances.}
	\label{tab:my-table2}
	\setlength{\tabcolsep}{3mm}{
		\begin{tabular}{ccc}
			\hline
			\textbf{Dataset}            & \textbf{Naturalness} & \textbf{Similarity} \\ \hline
			VCTK(3F,3M)\cite{veaux2016superseded}     & 3.53$\pm$0.08       &  3.82$\pm$0.07        \\
			LibriSpeech(2F,2M)\cite{panayotov2015librispeech}             & 3.12$\pm$0.10        & 3.51$\pm$0.08        \\
			LJSpeech(1F)\cite{ito2017lj}    &2.82$\pm$0.17        &3.05$\pm$0.12        \\
			VCC2020(2F,2M)\cite{zhao2020voice}     & 3.22$\pm$0.08           & 3.63$\pm$0.08        \\ \hline
		\end{tabular}
	}
\end{table}
\subsection{MediumVC} MediumVC performs A2A VC with utterance-wise optimization. The architecture shown in Fig. 3 is made of four modules: speaker encoder (\emph{E$_{s}$}($\cdot$)), SingleVC (\emph{V$_{Y}$}($\cdot$), pre-trained for speaker \emph{Y}), content encoder (\emph{E$_{c}$}($\cdot$)) and Decoder (\emph{D}($\cdot$,$\cdot$)). The structural units of \emph{E$_{c}$}($\cdot$) and \emph{D}($\cdot$,$\cdot$) are basically consistent with SingleVC, except for expanding the depths and widths of the networks to accommodate the representations of  multiple speakers. Additionally, the HiFi-GAN\cite{kong2020hifi} is introduced as vocoder. The conversion process can be noted as Eq. 5,6, and to minimize the  reconstruction loss as Eq. 7.
\begin{align}
\hat{Y_{i}}&=V_{Y}(X_{i}) \\
\hat{X_{i}}&=D(E_{s}(X_{i}),E_{c}(\hat{Y_{i}})) \\
\begin{matrix}
\textbf{min}\\ 
E_{s}(\cdot ),D_{s}(\cdot,\cdot )
\end{matrix}&=E[\begin{Vmatrix}
X_{i}-\hat{X}_{i}
\end{Vmatrix}_{1}^{1}]
\end{align}\\
The key of MediumVC is to build an asymmetrical reconstruction task ($\hat{\emph{Y$_{i}$}}$ $\to$ \emph{X$_{i}$}) by  processing \emph{X$_{i}$} $\to$ $\hat{\emph{Y$_{i}$}}$ in advance using \emph{V$_{Y}$}($\cdot$). For \emph{E$_{c}$}($\cdot$), it only deals with converted Y-related speeches. Instead of dealing with multiple speakers, it seems more efficient to extract specific speaker-independent features. Moreover, \emph{D}($\cdot$,$\cdot$) pay more attention on speaker embeddings $E_{s}(X_{i})$ to synthesis $\hat{X_{i}}$ because  there normally is no X-related information coming from $E_{c}(\hat{Y}_{i})$. Therefore, the implement of SSIF ensures sufficient feature decoupling and fusion of MediumVC, which achieves superior robustness to unseen speakers. Meanwhile, the timbre similarity has also been improved.
\section{Implement details}
\label{sec:pagestyle}
\begin{figure}[t]
	\centering
	\includegraphics[width=1.0\linewidth]{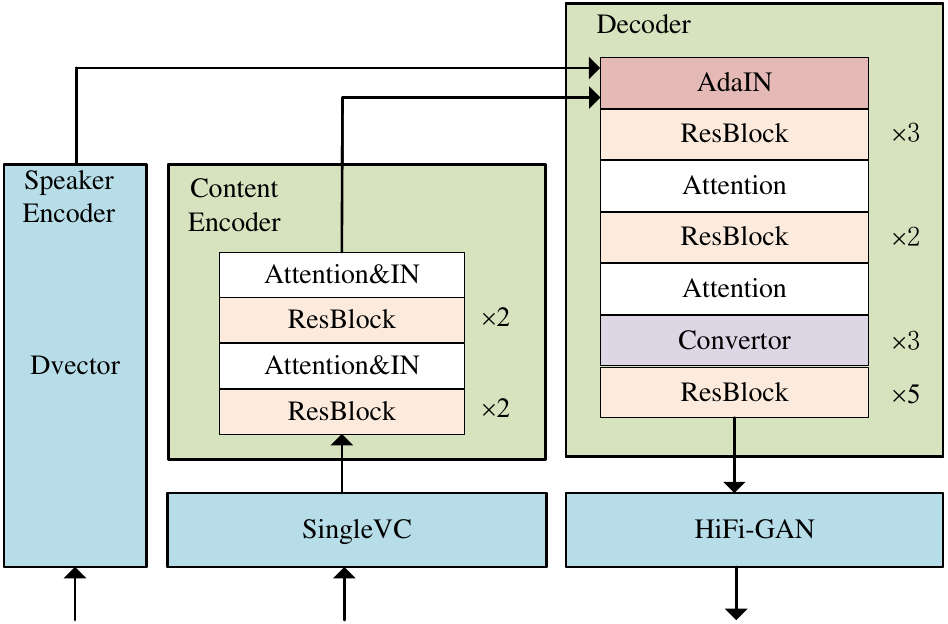}
	\caption{The architecture of MediumVC.}
	\label{fig:match}
\end{figure}
\subsection{Speaker Encoder}
Dvector\footnote{https://github.com/yistLin/dvector}  is a robust  speaker verification (SV) system pre-trained on VoxCeleb1\cite{Nagrani17}  using GE2E loss\cite{wan2018generalized}, and it  produces 256-dim speaker embedding. In our evaluation on multiple datasets (VCTK with 30000 pairs, Librispeech with 30000 pairs and VCC2020 with 10000 pairs), the equal error rates (EERs) and thresholds (THRs) are recorded in Table 3. Additionally, Dvector is also employed to calculate SV accuracy (ACC) of pairs produced by MediumVC for objective evaluation in Section 4.
\begin{table}[]
\centering
\caption{Evaluation on Dvector}
\label{tab:Dvector Evaluatio}
\begin{tabular}{cccc}
\hline
\textbf{Dataset} & VCTK & LibriSpeech   & VCC2020    \\ \hline
\textbf{EER(\%)/THR}    & 7.71/0.462    & 7.95/0.337 & 1.06/0.432 \\ \hline
\end{tabular}
\end{table}

\subsection{Content Encoder}
Speeches from multi-speakers are first processed by \emph{V$_{Y}$}($\cdot$) to synthesis Y-related speeches, which are used as inputs of \emph{E$_{c}$}($\cdot$). Therefore, to extract content information, \emph{E$_{s}$}($\cdot$) just need to deal with specific Y-related information, which greatly simplifies the performance requirements of \emph{E$_{c}$}($\cdot$). The inputs are first encoded as  36-dim bottleneck features and then are expanded to 256-dim representations to accommodate the speaker embeddings.

\subsection{Decoder}
Generally, speaker-related information is considering as global information, which remains consistent in any segments of $X_{i}$. \emph{D}($\cdot$,$\cdot$) accepts speaker embeddings $E_{s}(X_{i})$ and content embeddings $E_{c}(\hat{Y}_{i})$ as inputs. To realize the  global impact of $E_{s}(X_{i})$ on $E_{c}(\hat{Y}_{i})$, \emph{AdaIN} layer is introduced to fuse the two parts. $E_{c}(\hat{Y}_{i})$ has been processed with \emph{IN}, then $E_{s}(X_{i})$  and its transformation $\sqrt{\mid{}E_{s}(X_{i})\mid{}}$are used as affine transformation parameters $\beta$,$\alpha$, shown in  Eq. 8,9.
\begin{align}
AdaIN(E_{c}(\hat{Y}_{i}),E_{s}(X_{i}))&=\mathbf{\alpha} * E_{c}(\hat{Y}_{i})+\mathbf{\beta} \\
\mathbf{\alpha}=\sqrt{\begin{vmatrix}
E_{s}(X_{i})
\end{vmatrix}} &,\ \mathbf{\beta}=E_{s}(X_{i})
\end{align}
Ideally, $E_{c}(\hat{Y}_{i})$ don't contain target X-related information, and $\emph{D}(\cdot,\cdot) $ will rely more on the $E_{s}(X_{i})$ for building X-related speeches. It can enhance the dominant role of speaker embeddings on phonetic timbre construction, ultimately leading to higher target timbre similarity. Furthermore, the asymmetrical reconstruction mode($\hat{\emph{Y$_{i}$}}$ $\to$ $\hat{\emph{X$_{i}$}}$) promotes  model to capture more robust features for unseen speakers.
\subsection{Vocoder}
The HiFi-GAN\cite{kong2020hifi} vocoder is employed to convert log mel-spectrograms to waveforms. The model is trained on universal datasets with 13.93M parameters. Through our evaluation, it can synthesize 22.05 kHz high-fidelity speeches over 4.0 MOS, even in cross-language or noisy environments.
\subsection{Training details}
The training is performed on the VCTK corpus, containing 44 hours of utterances from 109 speakers, and each utterance is resampled to 22.05kHz. To perform utterance-wise optimization, MediumVC accepts speeches with variable lengths as inputs, and pre-processed features need padding to the same length for a mine-batch. We employ a two-step normalization scheme. First, to eliminate the effects of irregular volumes and inconsistent background noises, speech samples are uniformly scaled to 0\textasciitilde0.95. Then, the extracted 80-dim mel-spectrums are normalized to 0\textasciitilde1 to promote training. MediumVC is optimized with the AdamW optimizer with a learning rate = 0.0001 and Exponential LR is employed for the learning rate schedule. We use L1 loss functions without any tricks, the model performs a steady convergence. The samples\footnote{https://brightgu.github.io/MediumVC/} and codes\footnote{https://github.com/BrightGu/MediumVC} are released online.


\begin{table*}[]
\centering
\caption{The evaluations on MediumVC and  contrast methods. Mw/uS means MediumVC with untrained SingleVC and Mw/oS means MediumVC without SingleVC. The MOS results are statistically calculated with 95\% confidence intervals.}
\label{MediumVC evaluations}
\begin{tabular}{cccccccc}
\hline
\textbf{Metric} & \textbf{Dataset} & MediumVC              & Mw/uS                 & Mw/oS                 & FragmentVC            & AutoVC                & AdaIn-VC \\ \hline
Params          & /               & 26.41M                     & 26.41M                     & 21.28M                     & 48.01M                     & 28.42M                     & 9.04M        \\ \hline
Naturalness     & VCTK            & \textbf{3.52$\pm$0.11}                    & 3.41$\pm$0.13                     & 3.27$\pm$0.13                     & 3.0$\pm$0.17                     & 2.67$\pm$0.20                     & 3.21$\pm$0.11        \\
                & LibriSpeech        &  \textbf{3.34$\pm$0.13}                     & 3.07$\pm$0.15                     & 3.04$\pm$0.13                     & 2.73$\pm$0.15                                         & 2.40$\pm$0.18                                         & 2.86$\pm$0.13        \\
                & VCC2020             & \textbf{3.52$\pm$0.21}                     & 3.33$\pm$0.23                    & 3.28$\pm$0.19                     & 2.93$\pm$0.24                     & 2.45$\pm$0.20                    & 3.00$\pm$0.23        \\ \hline
Similarity      & VCTK            &  \textbf{3.82$\pm$0.21}                     & 3.80$\pm$0.17                     & 3.67$\pm$0.19                     & 3.22$\pm$0.19                    & 3.05$\pm$0.24                                        & 3.52$\pm$0.14        \\
                & LibriSpeech        & 3.25$\pm$0.16                     & 3.21$\pm$0.16                     & 3.17$\pm$0.21                     & 3.07$\pm$0.18                     & 2.96$\pm$0.20                     &  \textbf{3.43$\pm$0.18}        \\
                & VCC2020             & 3.36$\pm$0.20                     & 3.12$\pm$0.31                     & 3.09$\pm$0.23                     & 3.15$\pm$0.29 & 3.02$\pm$0.21                     &  \textbf{3.52$\pm$0.18}        \\ \hline
ACC(\%)             & VCTK            & 99.50                     & 92.16                     & 90.79                     & 92.35                     & 39.00                     &  \textbf{99.65}        \\
                & LibriSpeech        & 93.15                     & 88.43                     & 90.76                     & 87.01                     & 16.00               &  \textbf{98.79}        \\
                & VCC2020             & 99.45                     & 90.20                     & 91.37                     & 95.13                     & 17.00                     &  \textbf{99.50}       \\ \hline
\end{tabular}
\end{table*}

\section{experiments and discussions}
\label{sec:typestyle}

\subsection{Evaluation metrics}
We conduct evaluations on multiple datasets. For subjective evaluation, converted speeches are randomly selected from VCTK(seen speakers, 5F and 5M), LibriSpeech(unseen speakers, 5F and 5M)  and VCC2020(unseen speakers, 2F and 2M) to conduct evaluations on naturalness and similarity by the mean opinion score (MOS) test, and each speaker provides 5 utterances. For objective evaluation, converted speeches combined with target speeches are paired to calculate SV accuracy(ACC) using Dvector in Section 3.1. For each method shown in Table 4, there are 2000 randomly converted speeches are used for objective evaluation, except 100 speeches for AutoVC.
\subsection{Ablation Study}
We carry out two ablation experiments: MediumVC with untrained SingleVC(Mw/uS) and MediumVC without SingleVC(Mw/oS). All experiments are conducting under the same conditions. The results in Table 4 present that, for seen speakers in VCTK, all three perform closely on similarity while MediumVC outperforms on naturalness with slight advantages. However, the differences are expanded both on similarity and naturalness when evaluated on unseen speakers. It indicates that, compared with encoder-decoder-based methods(Mw/uS, Mw/oS), pre-trained SingleVC plays an important role in improving the robustness of MediumVC. Moreover, the objective evaluation on more converted speeches further suggests that MediumVC is rather robust in similarity. Besides, this robustness in timbre construction is not derived from a pre-trained speaker encoder but sufficient feature decoupling and fusion. We consider the implementation of SSIF helps to build asymmetrical reconstruction tasks, which eventually leads to outstanding performances,  especially promotes the global effects of speaker embeddings on the content information.
\subsection{Comparison with A2A VC}
We compare MediumVC with three A2A VC methods: FragmentVC\cite{lin2021fragmentvc}, AutoVC\cite{qian2019autovc} and AdaIN-VC\cite{chou2019one}. The contrast experiments are all conducted with the latest released pre-training model trained on VCTK. The results in Table 4 demonstrate that, for seen or unseen speakers, MediumVC performs better both in naturalness and similarity. The advance in naturalness indicates, compared with Wav2Vec 2.0-based embeddings in FragmentVC and deep content-related features in AutoVC and AdaIN-VC, SSIF maintains more robust speaker-independent features. Meanwhile, removing the influence of multi-speakers and building a specific-speaker periodic pattern further promote the advance in similarity. Additionally, AdaIn-VC with the least parameters achieves the best performances on ACC. We consider the main problem is the employing of discrete speaker embeddings produced by extra pre-trained SV systems in the other four methods(except FragmentVC). Compared to discrete speaker embeddings, that from the Autoencoder-based model(AdaIn-VC) seems to be smoother and more adaptable to unseen speakers. Experiments indicate that there are still plentiful noteworthy details in the research of robust A2A VC, such as the framework of models, the representation of the phonetic features \emph{etc}.

\section{Conclusion}
\label{sec:majhead}
In this paper, we propose SingleVC to perform A2O VC, and based on it, we propose MediumVC to perform A2A VC. The key to well-performance is that we build asymmetric reconstruction tasks for self-supervised learning. For SingleVC, we employ PSDR to edit source pitches, promoting the SingleVC to learn robust content information by rebuilding source speech. For MediumVC, employing SSIF processed by SingleVC promotes MediumVC to rely more on speaker embeddings to enhance target similarity. It is asymmetric tasks that drive models to learn more robust features purposefully.



\small
\bibliographystyle{IEEEbib}

\bibliography{icassp_mediumvc}

\end{document}